%% file: qfb5.tex
\documentclass[letterpaper,10pt,conference]{ieeeconf}

\IEEEoverridecommandlockouts
\usepackage{times}
\usepackage[dvips]{color}
\usepackage{amssymb}
\usepackage{amsmath}

\usepackage{epsfig}
\usepackage{cite}
\usepackage{verbatim}
\usepackage{subfigure}
\usepackage{multicol}
\usepackage{pstricks}
\usepackage{pst-node}

\newtheorem{theorem}{Theorem}
\newtheorem{lemma}{Lemma}

\title{\LARGE \bf
  Wireless Erasure Networks With Feedback}
\author{Brian Smith and Babak Hassibi%
\thanks{The first author would like to acknowledge funding from an ARO Young Investigator Award.  The work of the second author was in parts supported by the National Science Foundation under grant CCF 0729203, by the David and Lucille Packard Foundation and by Caltech's Lee Center for Advanced Networking.}%
\thanks{The authors are with the Department of Electrical and Computer Engineering, The University of Texas at Austin, Austin, TX; and with the Department of Electrical Engineering, The California Institute of Technology.  {\tt\small bsmith@ece.utexas.edu, hassibi@systems.caltech.edu}}%
}

\begin{document}

\onecolumn

\maketitle
\thispagestyle{empty}

\begin{abstract}

Consider a lossy packet network of queues, communicating over a wireless medium.  This paper presents a throughput-optimal transmission strategy for a unicast network when feedback is available, which has the following advantages:  It requires a very limited form of acknowledgment feedback.  It is completely distributed, and independent of the network topology.  Finally, communication at the information theoretic cut-set rate requires no network coding and no rateless coding on the packets.  This simple strategy consists of each node randomly choosing a packet from its buffer to transmit at each opportunity.  However, the packet is only deleted from a node's buffer once it has been successfully received by the final destination.  

\end{abstract}


\section{Introduction}
\label{sec-introduction}
It is well know that in the point-to-point channel model, feedback can never increase the value of the information theoretic capacity\cite{CTtext}.  However, there several significant advantages to having feedback.  Feedback allows coding strategies which can significantly increase the probability of error exponent, for example the Schalkwijk-Kailath scheme for additive Gaussian noise channels\cite{SK}.  Feedback can also allow transmission strategies with extremely simple coding algorithms.  Specifically, consider the binary symmetric erasure channel.  When feedback is available, the transmitter can simply repeat each bit until successfully received.  Capacity is achieved, and in some sense, no coding whatsoever is required.  

In this paper, a unicast model of a lossy wireless network of queues is considered, similar in spirit to the wireless erasure network\cite{DGH:04,DGH:06}.  Our network model is characterized by independent erasure channels/loss probabilities on a directed graph, a wireless broadcast requirement, asynchronous transmission timing, and unicast for a single source-destination pair.  With transmit opportunities occurring as a unit rate Poisson process, a transmission by one node will be received independently with some fixed probability by each other node in the network.  The network model will allow general feedback, but it will be shown that only a very limited form of acknowledgment feedback is required to achieve the throughput-optimal cut-set capacity.  The primary differences between our model and that of \cite{DGH:06} are first, the availability of feedback, and second, an asynchronous, memoryless arrival process (rather than a slotted-time model).  The authors do not believe the difference in timing model to be critical, and conjecture that the given transmission algorithm will be throughput-optimal in a slotted-time model.  Extensive simulation on the slotted-time model has inspired confidence in this conjecture.  Additionally, the addition of feedback eliminates the requirement for any side-information concerning the location of erasures throughout the network to achieve capacity, in contrast with the decoding strategy of \cite{DGH:06}.  

A similar asynchronous network model was studied in \cite{Lun}.  The authors' model demonstrates the usefulness of network coding:  with no feedback, but allowing network coding and additionally, a packet header describing the linear combinations of data packets included in the transmission, they demonstrate the achievability of the cut-set bound.  Our work highlights somewhat of a dual statement:  without any sort of coding, but with feedback, the same cut-set packet rate is achievable.

The paper \cite{Neely} also is concerned with a similar wireless lossy packet network model.  With a backpressure algorithm, throughput-optimality in a multi-commodity sense is also achieved in a multiple-source multiple-destination network.   This algorithm requires link-level feedback, and for each node to maintain knowledge of the queue state of, in worst case, every other node in the network.  It provides a decision process, when multiple nodes in the network receive copies of the same packet, to determine which (if any) of those nodes should keep that packet and attempt to forward it onward.

In contrast, the routing algorithm described in this paper is completely decentralized and requires no conferencing among nodes to decide who should ``keep'' a packet that it has received.  Instead, there will in general be multiple copies of each packet throughout the network.

Specifically, the algorithm is as follows:  Whenever a node has an opportunity to transmit a packet, it will randomly choose one packet from its buffer.  Every time that a packet successfully reaches the final destination node, that node will (errorlessly) broadcast an acknowledgment to every node in the system stating that this particular packet has successfully completed its transit of the network.  Only after receiving this acknowledgment from the \textit{final destination node} will any node remove the packet from its buffer.  Indeed, the entire network will then flush that packet from all the buffers.  This paper shows via Foster's Theorem and an application of an appropriate and novel Lyapunov function the stability of all network queues under this operation as long as the input data rate is less than the minimum-cut of the network.  The authors are unaware of previous uses of an exponential Lyapunov function of the form we consider in showing stability results.

The advantages of this throughput optimal strategy include
\begin{itemize}
\item It requires no coding, particularly no network coding at intermediate nodes.
\item The only information that a packet header must contain is an identifier - no additional information is required.
\item It is completely decentralized.  No coordination or conferencing, other than the acknowledgment feedback, is required.
\item It is topology independent.  No node other than the source needs any information about the layout of the network.  The source must only be given the value of the min-cut, which could even be adaptively estimated, if desired.
\item The only feedback required, a simple acknowledgment from the destination, is practically already implemented in real systems.
\end{itemize}
The main thrust of this paper:  A demonstration that, in this lossy wireless network, feedback obviates the need for coding, network coding in particular.

\section{Network Model and Notation}
\label{sec:sysmodel}

Consider a directed (possibly cyclic) graph $G(V,E)$ with $n+2$ nodes: a source node, a destination node, and $n$ intermediate nodes.  Label the source node $s$, the destination $d$, and index the other nodes as $i\in{1,..,n}$.  To each edge pair $(i,j)\in V \times V$ assign an erasure probability $0\le \epsilon_{ij} \le 1$.  If the directed edge $(i,j)$ does not exist in the graph, then assign $\epsilon_{ij}=1$.  Define $\mu_{ij}=1-\epsilon_{ij}$.

Because of the wireless nature of the model, when a node $i$ transmits a packet, each other node in the system $j$ has the probability $\mu_{ij}$ of successfully receiving that packet.  The events that packets are dropped are independent, that is i.i.d. across time for any fixed edge $(i,j)$, and independent between every pair of edges.  We will consider the case where the events corresponding to combinations of packet drops from a single transmitter at a fixed time can be correlated in a later section. 

Allow an infinite buffer to exist at each node in the network.  Packets will exogenously arrive at the source node $s$ according to a Poisson process with arrival rate $\lambda$.  At average rate $1$ exponentially distributed intervals, each node in the network (other than the destination node) receives an opportunity to transmit a packet.  

Each packet has a unique identifier in its header.  Therefore, if a node already has a copy of a particular packet and it receives that packet again, the contents of that node's buffer remain unchanged. 

A feedback mechanism exists such that when the destination node receives a packet, it instantaneously, via a delay-free feedback, notifies all of the other nodes in the system of that fact.  All nodes in the system can then immediately remove that particular packet from their buffer.  

Finally, this asynchronous model does not consider any receiver interference or the possibility of simultaneous arrivals.  It is, however, possible to take into account interference or collisions by appropriately assigning the erasure probabilities of the model, accounting for lost packets by increasing the probability of packet drops.

\section{Cut-set Upper Bound and Transmission Strategy}

Under any transmission strategy, the cut-set upper-bound remains valid.  Intuitively, the cut-set upper-bound is obtained by dividing the network into two parts $S$ and $S^C$ and creating two super-nodes.  That is, by allowing free, unlimited communication among the nodes in $S$ and among the nodes in $S^C$, we can only increase the capacity of the system.

With that in mind, let $S$ be a subset of the $n+2$ nodes such that $s\in S$ and $d\in S^C$.  There are $2^n$ such subsets.  Let $\mathcal{S}$ be the set of all such subsets.  The super-node created by joining all nodes in $S$ together will still have opportunities to transmit at exponentially distributed intervals, but now the sum rate will be $|S|$ -- a rate of $1$ for each node in $S$. For each node $i\in S$, because of the unlimited free communication on the right side of the cut in $S^C$, as long as one of the nodes $j\in S^C$ successfully receives the packet, we can count it in the total communication throughput.  Therefore, define
\begin{equation}
C(S)=\sum_{i\in S} \left ( 1 - \prod_{j\in S^C} \epsilon_{ij} \right )
\end{equation}
as the cut-set capacity for the subset $S$, i.e. an upperbound on the rate of packets that can be transmitted across the $S-S^C$ cut, exactly as per \cite{DGH:06}.  Note that this upperbound is valid whether or not there is feedback in the system.  This is because the procedure for obtaining the cut-capacity reduces the network to a memoryless point-to-point channel, for which feedback does not increase capacity \cite{CTtext}. 

For convenience sake, define 
\begin{align} \label{eq:ci}
C_i(S) = 1-\prod_{j\notin S} \epsilon_{ij} 
\end{align}
for pairs $\left(i,S\right)$ such that  $i\in S$.  $C_i(S)$ represents the contribution to the cut-set bound for the cut $S$ from the node $i\in S$. 

The total throughput $T<C(S)$ then, for every subset $S$, and
\begin{equation*}
R<\min_{S\in \mathcal{S}} C(S). 
\end{equation*}
In \cite{DGH:06}, the authors show that, with the packet erasure locations known at the destination and the appropriate use of network coding, this min-cut capacity is indeed achievable in a wireless erasure network.

The authors would like to emphasize the key role that the subsets $S$ will play in the proof and the derivation of the stability results.   The minimum of $C(S)$ over all $S-S^C$ cuts must emerge from any stability equations; therefore it is reasonable that each cut-set represented by $S$ must play a role.  As will be further explained, the sets $S$ will become essential as indices to the variables $m_S$ which describe the state of our Markov chain model.  It will become clear that as the state variable $m_S$ corresponding to the subset $S$ becomes large, the requirement $\lambda < C(S)$ becomes a dominant constraint. 

The network operates in the following manner:  At every transmission opportunity for a node, that node \textit{randomly} chooses one of the packets in its buffer to transmit.  If the buffer is empty, then that transmission opportunity is lost.  Only when acknowledgment from the final destination $d$ is received will a node remove a packet from its buffer; therefore in general there are multiple copies of each packet in the network.

\begin{theorem}\label{th:1} 
Under this randomized transmission strategy, all queues in a wireless erasure network with feedback are stable as long as $\lambda<C(S)$ for all $S\in \mathcal{S}$. 
\end{theorem}

At first glance, this randomized strategy seems unnecessarily wasteful.  Consider a network which is a simple serial line of queues.  In this case, it is obvious that an optimal strategy, when link-level feedback is available, is to stop attempting to transmit a packet (and remove it from one's queue) as soon as it is successfully received at the next queue down the line.  Leaving a successfully transmitted packet in the queue could result in the retransmission of that packet, possibly wasting a transmission opportunity that could be put to better use sending a new packet.  

However, the randomization is crucially important in achieving the minimum-cut value for this network and for a general network.  To achieve the min-cut, it is essential that all transmitters on the min-cut boundary transmit packets at almost every channel use and that these packets be almost always distinct.  As the input rate $\lambda$ increases, the min-cut slowly becomes the bottleneck of the network and the queues on its boundary will grow large.  This will ensure that each transmitter always has a packet to transmit with high probability.  The randomization in packet transmission guarantees that for such long queues the probability that two transmitters along the min-cut transmit the same packet is very low.  Deterministic strategies, such as FIFO for example, cannot guarantee this without coordination, and so the randomized strategy is essential to achieving the optimal throughput in a completely decentralized manner.

In the line network in particular, edges which are not the minimum cut can afford to retransmit a certain number of packets, since they have extra capacity.  In fact, edges which lie downstream of the minimum cut edge will have relatively short queue lengths (compared to the queues upstream of the minimum cut edge) since they can remove packets from their queue at a faster rate than those packets can arrive across the minimum cut edge.

All queues upstream of the minimum cut, however, will have a relatively large number of packets.  If many packets are transmitted multiple times across the minimum-cut edge, then the queue length at that edge will grow large.  However, as the queue length grows large (with new arrivals), the probability of picking a ``useless'' packet will \textit{decrease} as most of the packets in the queue have not yet been successfully sent. This unwanted probability will be made as arbitrarily small as required (depending on the ratio between $\lambda$ and the minimum $\mu$) as the queue length grows.  

Note that strategies such as the one in \cite{Neely} implement an algorithm to assure that there is only one copy of each packet in the network at a time.  Such strategies necessarily require some amount of link-level feedback and inter-node communication to guarantee the single copy, under the broadcast nature of the wireless medium.  The strategy of this paper eliminates the need for any additional intra-network communication, other than the single feedback acknowledgment.

\section{Proof Preliminaries}
\subsection{Notation and Description of Markov Chain Model}\label{sec:MCD}
Before formally beginning the proof of Theorem \ref{th:1}, some additional notation must be defined.

The subset $S$ has already been defined to be an element of $\mathcal{S}$, which is essentially the power-set of $n$.  Precisely, $\mathcal{S}$ differs the power-set of $n$ only in that all $S\in \mathcal{S}$ always include the source node $s$ and never include destination node $d$.  Equivalently, each element $S$ can represent an index in the set $\{0,1,2,...,2^n-1\}$.  With this notion, the length-$n$ binary expansion of $S$ indicates which of the $n$ nodes are contained within the subset $S$.  This yields a one-to-one correspondence between subsets, cut-sets, and indices, all represented by the overloaded notation $S$.  

A continuous time Markov chain model is used to describe the state of the queuing network.  Transitions between states will occur when one of three different types of events happen in the network:
\begin{itemize}
\item A new packet is received (at rate $\lambda$) by the source node $s$.
\item A packet is successfully transmitted from some node $i$ in the system to some subset of the receivers.
\item A packet is successfully received by the the destination node $d$ and therefore exits the network.
\end{itemize}
By the asynchronous, continuous time model of the network, no two of these events can occur simultaneously.   

In the $n=1$ three node network, the size of the buffers at the source node $s$ and the intermediate node $1$ are sufficient to describe any state of the system.  There must be more packets at the source node $s$ than at the intermediate node $1$ at any point in time.  By the given network operation protocol, no packet is deleted from a queue until it reaches the final destination, so that if a packet is present anywhere within the system, it must be present at the source node $s$.  

One option for the state variable of the system is to use $\underline{q}=(q(s),q(1))$, representing the lengths of the queues at source and relay nodes respectively.  This notation has the disadvantage that there would exist constraints such as ``the number of packets at $1$ must be smaller than the number of packets at $s$'', i.e. $q(s)\ge q(1)$, on the state space.  In addition, when a network contains more than a single intermediate node, knowing the queue lengths alone is not sufficient to describe the state of a system where multiple copies of each packet may occur at different nodes.  When a packet leaves the network, the model must be able to determine at exactly which nodes in a system the queue lengths should be decreased.  To completely describe the system state and to eliminate the need for any such constraints on state variables, we consider an alternate notation.

Let $m_1$ be the number of packets which appear at both the nodes $s$ and $1$.  Let $m_0$ be the number of packets which appear at the source node uniquely.  Then the source node has a total of $m_0 + m_1$ packets, while the relay node has exactly $m_1$ packets in its buffer.

This state description can be generalized to an $n+2$ node network.  The Markov chain describing the system state is a vector $\underline{m}$ with $2^n$ dimensions:
\begin{equation}
\underline{m}=\left(m_0,m_1,...,m_S,...,m_{2^n-1}\right)
\end{equation}

The dimensions of the state vector $\underline{m}$ are indexed by the subsets $S\in\mathcal{S}$.  The value $m_S$ is the number of packets which appear at every node $i\in S$ and at no node $j\in S^C$.  Therefore, the number of packets $q(i)$ which appear any node $i\ne s,d$ in the network is a function of $\underline{m}$.  Let 
\begin{equation*}
\mathcal{S}_i = \left\{S\in\mathcal{S}|\text{the $i^{th}$-least significant bit in the binary expansion of $S$ is a $1$}\right\}.
\end{equation*}
Then 
\begin{equation*}
q(i) = \sum_{S\in\mathcal{S}_i} m_S,
\end{equation*}
while the destination node $d$ retains no buffer, and the source node $s$ has
\begin{equation*} 
q(s) = \sum_{S\in\mathcal{S}} m_S
\end{equation*}
packets in its buffer.

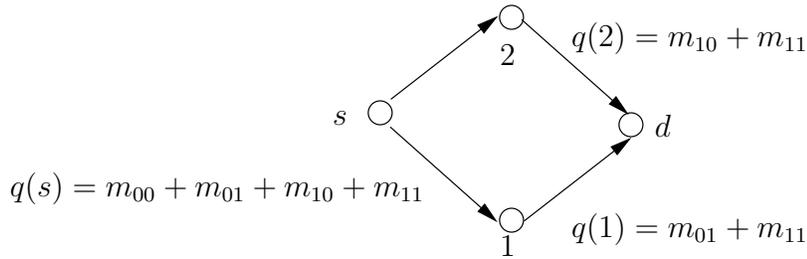
\begin{figure}[ht]
\begin{center}
\input{2node.pstex_t}
\end{center}
 \caption{Relationship between queue lengths and $\underline{m}$ for the case $n=2$} \label{fig:ex_2node}
\end{figure}

Figure \ref{fig:ex_2node} illustrates the queue lengths for a network with $n=2$, using the binary expansion of the $S$ indices.

\subsection{Markov Chain Evolution - Transition Model} 

To understand the evolution of the Markov chain model describing the state $\underline{m}$ of the queuing system, first take an example of the network where $n=1$.  

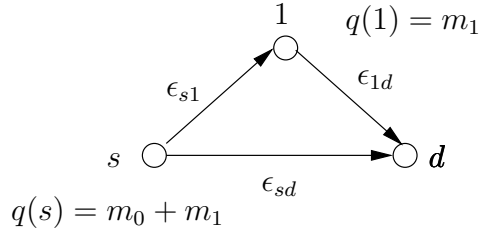
\begin{figure}[ht]
\begin{center}
\input{1node.pstex_t}
\end{center}
 \caption{A general $n=1$ wireless erasure network.} \label{fig:ex_1node}
\end{figure}

Successful transmission events can cause four different kinds of transitions to the state vector $\underline{m}=\left(m_0,m_1\right)$.  
\begin{itemize}
\item There is an exogenous arrival to the system.  In this case, the source node receives a new packet; the source is therefore the only node in the system which has that particular packet in its buffer.  Thus, the value of $m_0$ is increased by $1$. 
\item A packet \textit{that is present exclusively at the source node $s$} can be successfully received by the destination $d$, and therefore flushed from the network.  Note that this is a subset of the event \textit{a packet transmitted from the source is received by the destination} - packets at the source can belong to either the subset $m_0$ (with probability $m_0/\left(m_0+m_1\right)$) or to the subset $m_1$.  The value of $m_0$ is decreased by $1$.
\item  A packet \textit{that is present at the relay node}, i.e. counted in the variable $m_1$, can be successfully received by the destination $d$, and therefore flushed from the network.  If this packet was transmitted by the source, it must have come from the set of $m_1$ packets (with probability $m_1/\left(m_0+m_1\right)$).  If this packet was transmitter by the relay node $1$, then by definition it must have been one of the $m_1$ packets at both the source and the relay.  In either case, the value of $m_1$ decreases by $1$. 
\item If the transmitter selects a packet from the set of $m_0$ packets, and that packet is successfully received by node $1$, but not by the receiver, then that particular packet would now be in both nodes' queues.  In that case, we have a transition in which $m_0$ decreases by $1$ (there is one less packet which is unique to node $1$) and $m_1$ increases by $1$ (there is one additional packet which is located at both the source node $s$ and the relay node $1$.)
\end{itemize}

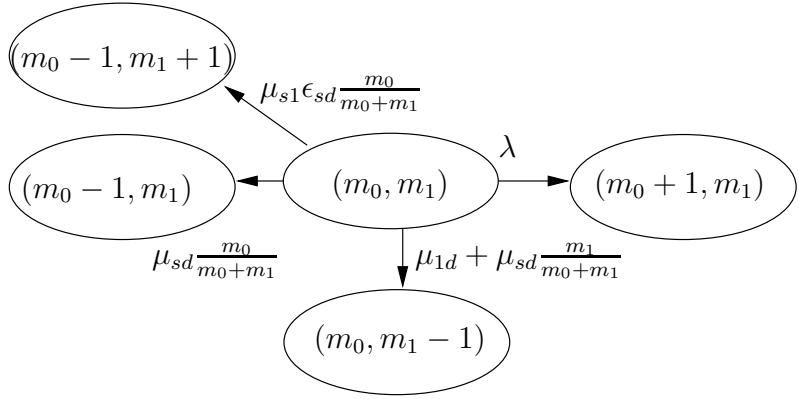
\begin{figure}[ht]
\begin{center}
\input{1node_chain.pstex_t}
\end{center}
 \caption{Possible transitions and transition rates from a state $(m_0,m_1)$ in the $n=1$ wireless erasure network.} \label{fig:1node_chain}
\end{figure}

Each of these possible transitions and their individual rates are illustrated in Figure \ref{fig:1node_chain}.

In general, a network with $n$ relay nodes has these same three kinds of transitions:
\begin{itemize}
\item A packet arrives at the source node, with rate $\lambda$.  In this case, $m_0$ increases by $1$.
\item A packet (which exists in the subset $S_1$ of nodes) exits the system from some node $i$, with rate 
\begin{equation*}
\sum_{i\in S_1} \mu_{id} \frac {m_{S_1}}{q(i)}.
\end{equation*}  
Here, $m_{S_1}$ decreases by $1$.
\item A packet (which exists in the subset $S_1$) transmitted at some node $i$ is successfully received at some subset of possible receiver nodes, at least one of which did not previously have that particular packet in its buffer.  In this case, let $S_2$ be the new subset of nodes which have this packet.  This constrains $S_1 \subset S_2$, and this occurs with rate
\begin{equation*}
\sum_{i\in S_1} \left(\prod_{j \in S_2/S_1}\mu_{ij} \prod_{j\notin S_2} \epsilon_{ij}\right) \frac{m_{S_1}}{q(i)}.
\end{equation*} 
Here, $m_{S_1}$ decreases by $1$ while $m_{S_2}$ increases by $1$.  It is important to note that in this kind of transition, the subset $S_2$ whose variable $m_{S_2}$ increases must always be a superset of the subset $S_1$ whose variable $m_{S_1}$ decreases.  
\end{itemize}

\subsection{Queue Stability and Foster's Theorem}
We desire to show that, for any arrival rate $\lambda < \min_{S\in \mathcal{S}} C(S)$, all the queues in the network are stable.  We associate stability with positive recurrence:  A state in a Markov chain is positive recurrent if the expected return time to that state is finite.  A Markov chain is positive recurrent if all states are positive recurrent.  We first present a review of Foster's Theorem, which is the main proof mechanism\cite{FCtext}.

\begin{theorem}{\textit{Foster's Theorem}.  }
Let the transition matrix $\textbf{P}$ on the countable state space $M$ be irreducible and suppose there exists a function $V:M\rightarrow \mathbb R$ such that $\inf_{ m} V( m) > -\infty$ and
\begin{align*}
\sum_{ k \in M} p_{m k} V(k)&< \infty &\text{ for all } m\in F \\
\sum_{ k \in M} p_{ m  k} V(k)&< V( m) - 1 &\text{ for all } m\notin F
\end{align*}
for some finite set $F$.  Then the corresponding homogeneous Markov chain is positive recurrent.
\end{theorem}

Intuitively, the theorem states that as long as there is a Lyapunov function which is on average decreasing, then the value of that function cannot go to infinity with increasing time. 
\section{Proof for the Case $n=1$}
\label{sec:n1}
This section contains a demonstration of the stability proof for the simplest network, the case where $n=1$, illustrated in Figure \ref{fig:ex_1node}.  Note that for this particular network, the cut-set bound evaluates to 
\begin{equation*}
 \min\left(1-\epsilon_{s1}\epsilon_{sd},1-\epsilon_{1d}+1-\epsilon_{sd}\right)
\end{equation*}

\begin{lemma}
The network illustrated in Figure \ref{fig:ex_1node} is stable for 
\begin{equation*}
\lambda < \frac{N}{N+1} \frac{1}{1+\delta} \min\left(1-\epsilon_{s1}\epsilon_{sd},1-\epsilon_{1d}+1-\epsilon_{sd}\right)
\end{equation*}
for any fixed $N>0$ and $\delta>0$.
\end{lemma}

By choosing $N>>1$ and $\delta<<1$ appropriately, for any $\lambda$ less than the cut-set bound, the randomized transmission policy with feedback stabilizes all the network queues.

\begin{proof}
Consider the Lyapunov function
\begin{equation}\label{eq:Lyap1}
V\left(m_0,m_1\right) = N \left(1+\delta\right)^{m_0} + \left(1+\delta\right)^{m_0+m_1}.
\end{equation}
This Lyapunov function is ``rewarded'' (i.e. decreases) when $m_0$ decreases and penalized when $m_0$ increases.  When a packet is received at the relay node, the function is rewarded (while $m_1$ increases, $m_0$ simultaneously decreases) and when a packet leaves the system (i.e. $m_1$ decreases) the function is also rewarded.

We identify three different cases to study.  These cases arise first, because several state transitions in the Markov chain of Figure \ref{fig:1node_chain} become unavailable in certain states (for example, when $m_0=0$ a packet cannot transition from the subset $m_0$ to the subset $m_1$).  Secondly, some of the cases individually give rise to the required cut-set constraints on $\lambda$ that the cut-set bound requires. 
\begin{itemize}
\item Case 1:  When $m_0=0$ and $m_1 > 0$.
\item Case 2:  When $m_0 > 0$ and $m_1 = 0$.
\item Case 3:  When both $m_0 >0$ and $m_1 > 0$.
\end{itemize}
As previously stated, when one of the variables in the state description $\underline{m}$ is equal to zero, one or more transitions from the state transition Figure \ref{fig:1node_chain} become unavailable.

\subsection{Case 1 : $m_0=0$}
Evaluate the expected change in the value of the Lyapunov function $V(0,m_1)$ to determine when it is bounded away from zero:
\begin{align*}
\lambda &\left(V\left(1,m_1\right)-V(0,m_1)\right) + (\mu_{1d}+\mu_{sd})\left(V(0,m_1-1)-V(0,m_1)\right)<0\\
\lambda &\left(N(1+\delta)^1 + (1+\delta)^{1+m_1} - N(1+\delta)^0- (1+\delta)^{m_1}  \right) \\
&+ (\mu_{1d}+\mu_{sd})\left( N(1+\delta)^0 + (1+\delta)^{m_1-1} - N(1+\delta)^0 - (1+\delta)^{m_1}   \right) <0\\
\lambda &\left ( N\delta + (1+\delta)^{m_1}(1+\delta-1)\right) + (\mu_{1d}+\mu_{sd})\left(1+\delta\right)^{m_1-1}\left(1-(1+\delta)\right)<0\\
\lambda &\left ( N + (1+\delta)^{m_1}\right) < (\mu_{1d}+\mu_{sd}) \left(1+\delta\right)^{m_1-1}\\
\lambda &< (\mu_{1d}+\mu_{sd}) \frac{(1+\delta)^{m_1-1}}{(1+\delta)^{m_1}+N}
\end{align*}
The first line represents the change in Lyapunov function for all possible state transitions from the state $\underline{m}=(0,m_1)$, weighted by the appropriate rates to calculate the expectation.  The right hand side of the final inequality approaches as arbitrarily close to $(\mu_{1d}+\mu_{sd}) \frac{1}{1+\delta}$ as desired for sufficiently large $m_1$.  That is, for any given $\lambda <  (\mu_{1d}+\mu_{sd}) \frac{1}{1+\delta}$, there exists a finite $\tilde{m}_1$ such that the expected value of the Lyapunov function decreases for states $\underline{m}=(0,m_1)$ for all $m_1>\tilde{m}_1$.

Note that for this case, one of the two cut-set bounds on $\lambda$ is obtained.

\subsection{Case 2 : $m_1=0$}
For $\underline{m}=(m_0,0)$, Figure \ref{fig:1node_chain} indicates that we desire

\begin{align*}
\lambda &\left(V\left(m_0+1,0\right)-V(m_0,0)\right) + \mu_{sd}\left(V(m_0-1,0)-V(m_0,0)\right)+\mu_{s1}\epsilon_{sd}\left(V(m_0-1,1)-V(m_0,0) \right)<0\\
\lambda &\left(N (1+\delta)^{m_0} +(1+\delta)^{m_0}\right) < \mu_{sd}\left(N(1+\delta)^{m_0-1}+(1+\delta)^{m_0-1}\right) + \mu_{s1}\epsilon_{sd} N(1+\delta)^{m_0-1} \\
\lambda &(N+1) (1+\delta) < \mu_{sd}\epsilon_{sd} (N+1) + \mu_{s1}\epsilon_{sd} N \\
\lambda &< \mu_{sd} \frac{1}{1+\delta} + \mu_{s1}\frac{N}{N+1}\frac{1}{1+\delta}
\end{align*}

As long as 
\begin{equation*}
\lambda < \left(1-\epsilon_{sd}+\epsilon_{sd}(1-\epsilon_{s1})\right)\frac{N}{N+1} \frac{1}{1+\delta} = \left(1-\epsilon_{sd}\epsilon_{s1}\right)\frac{N}{N+1} \frac{1}{1+\delta}
\end{equation*}
then the expected value of the Lyapunov function decreases for all states of the form $\underline{m}=(m_0,0)$.

\subsection{Case 3 : $m_0,m_1>0$}
All of the transitions in Figure \ref{fig:1node_chain} are possible from the state $\underline{m}=(m_0,m_1)$.

\begin{align*}
\lambda &(V(m_0+1,m_1)-V(m_0,m_1)) \\
&+\left(\mu_{1d}+\mu_{sd}\frac{m_1}{m_0+m_1}\right) (V(m_0,m_1-1)-V(m_0,m_1))\\
&+\left(\mu_{sd}\frac{m_0}{m_0+m_1}\right) (V(m_0-1,m_1)-V(m_0,m_1))\\
&+\left(\mu_{s1}\epsilon_{sd}\frac{m_0}{m_0+m_1}\right)(V(m_0-1,m_1+1)-V(m_0,m_1)) <0\\
\lambda &N(1+\delta)^{m_0} + (1+\delta)^{m_0+m_1}\\
&<\left(\mu_{1d}+\mu_{sd}\right)(1+\delta)^{m_0+m_1-1}\\
&+\mu_{sd}\frac{m_0}{m_0+m_1}N(1+\delta)^{m_0-1}+(1+\delta)^{m_0+m_1-1}\\
&+\mu_{s1}\epsilon_{sd}\frac{m_0}{m_0+m_1}N(1+\delta)^{m_0-1}\\
\lambda &\left(N(1+\delta) + (1+\delta)^{m_1+1}\right)\\
&<\left(\mu_{1d}+\mu_{sd}\right)\left(1+\delta\right)^{m_1}+\left(1-\epsilon_{s1}\epsilon_{sd}\right)\frac{m_0}{m_0+m_1}N
\end{align*}

By inspection, note that regardless of the value of $m_0$, an $m_1^*$ can be chosen sufficiently large such that if $\lambda < \left(\mu_{s1}+\mu_{sd}\right)\frac{1}{1+\delta}$, the expected value of the Lyapunov function is decreasing for all states with $m_1>m_1^*$.  Likewise, for any fixed $m_1$,  choose $m_0 > Nm_1$, and if $\lambda < \min(1-\epsilon_{s1}\epsilon_{sd},\mu_{1d}+\mu_{sd})$, then the expected value of the Lyapunov function is decreasing.  Thus, there are only a finite number of states where the expected value of Lyapunov function is increasing, and the requirements of Foster's Theorem are fulfilled.
\end{proof}


\section{Proof for General Network}
Recall Theorem \ref{th:1}, which we desire to prove: 

\textit{Theorem 1:}  Under the given randomized transmission strategy, all queues in a wireless erasure network with feedback are stable as long as $\lambda<C(S)$ for all $S\in \mathcal{S}$. 

For a general wireless erasure network with $n+2$ nodes, recall the Markov chain describing the system evolution described in Section \ref{sec:MCD}.  Foster's Theorem is utilized to demonstrate the stability of this Markov chain for a general $n+2$ node network.

\subsection{General Lyapunov Function}
For a $n+2$ node network, define the Lyapunov function $V(\underline{m})$ as

\begin{equation}\label{eq:L}
V\left(\underline{m}\right) = \sum_{S\in\mathcal{S}}{N_{|S|}\left(1+\delta\right)^{\sum_{S'\subseteq S} m_{S'}}}
\end{equation}
where the $N_{|S|}$ and $\delta$ are fixed constants.  The $N_{|S|}$ should jointly satisfy 
\begin{equation} \label{eq:define_n}
N_{|S|}>N\sum_{S'\supset S} N_{|S'|}.
\end{equation}

To form some intuition on the particular choice of Equation (\ref{eq:L}), consider the three kinds of transitions that can occur in our Markov Chain.  When a packet arrives in the system, $m_0$ increases by $1$, and the value of the Lyapunov function increases.  Whenever any other transition occurs, the system is, in some sense, advancing a packet towards the final destination, and we would like the value of the Lyapunov function to decrease.  This can happen in two ways:  
\begin{itemize}
\item A packet which appears in the subset $S_1$ of nodes can exit the system.  Then, $m_{S_1}$ will decrease, and all of the terms in the summation corresponding to $S\supseteq S_1$ will decrease in value.  (i.e., those terms which contain the factor $(1+\delta)^{m_{S_1}}$).  The Lyapunov function therefore decreases in value.
\item A packet which appears in the subset $S_1$ of nodes will arrive at some other nodes, and then will appear in the subset $S_2\supset S_1$.  Then, all the terms in the summation corresponding to $S$ such that $S_1\subseteq S$, but $S_2 \nsubseteq S$, will decrease in value (i.e. those which contain the factor $(1+\delta)^{m_{S_1}}$ but not $(1+\delta)^{m_{S_2}}$).  However, those which contain both the factor $(1+\delta)^{m_{S_1}}$ and $(1+\delta)^{m_{S_2}}$ will remain unchanged (since $m_{S_1}$ decreases by $1$ and $m_{S_2}$ increases by $1$).  Note that, since $S_1 \subset S_2$, whenever $m_{S_2}$ appears in an exponent, so does $m_{S_1}$.  Thus, $m_{S_2}$ never appears in isolation and none of the terms in the Lyapunov function increase.
\end{itemize}

\subsection{Proof}
The proof of Theorem \ref{th:1} follows directly from the following lemma:
\begin{lemma}
The expected value of the function $V(\underline{m})$, defined in Equation (\ref{eq:L}), is increasing only on a finite number of states whenever $\lambda < \frac{N}{N+1}\frac{1}{\left(1+\delta\right)^2}\min_{S\in\mathcal{S}} C(S)$.
\end{lemma}
Thus, for any $\lambda < \min_{S\in\mathcal{S}} C(S)$, we can find an appropriate Lyapunov function to show the system's stability by choosing $N$ sufficiently large and $\delta$ sufficiently small.

\begin{proof}

First fix $S\in\mathcal{S}$, and examine the term in the main summation of Equation (\ref{eq:L}) corresponding that $S$.   Then,  determine which transitions of the Markov chain effect the value of that term. 

Let 
\begin{equation}
V_S(\underline{m}) = N_{|S|} \left(1+\delta\right)^{\sum_{S'\subseteq S} m_{S'}}.
\end{equation}

An arrival to the system effects every $V_S$, since every term $V_S$ contains $m_0$.  Thus $\forall S\in\mathcal{S}$,
\begin{align}  \notag
V_S&(m_0+1,m_1,...)-V_S(m_0,m_1,...)\\ \notag 
&=N_{|S|}\left(1+\delta\right)^{\sum_{S'\subseteq S} m_{S'}+1} -  
N_{|S|}\left(1+\delta\right)^{\sum_{S'\subseteq S} m_{S'}}\\
&=\delta N_{|S|}\left(1+\delta\right)^{\sum_{S'\subseteq S} m_{S'}}
\end{align}
These events occur at rate $\lambda$.

If a packet appearing in cut-set $S_1$ departs the system, precisely the terms $V_S(\underline{m})$ when $S\supseteq S_1$ will decrease, since they are the only terms in the Lyapunov function Equation (\ref{eq:L}) which contain $m_{S_1}$.  For $S\supseteq S_1$, 
\begin{align} \notag
V_S&(m_0,m_1,...,m_{S_1}-1,...)-V_S(m_0,m_1,...)\\ \notag
&=N_{|S|}\left(1+\delta\right)^{\sum_{S'\subseteq S} m_{S'}-1}-
N_{|S|}\left(1+\delta\right)^{\sum_{S'\subseteq S} m_{S'}}\\
&=-\delta N_{|S|}\left(1+\delta\right)^{\sum_{S'\subseteq S} m_{S'}-1}.
\end{align}
These events will occur when any node $i\in S_1$ transmits a packet in $S_1$ which is successfully received by the destination node $d$.  Given an opportunity to transmit, the node $i$ chooses a packet in $S_1$ with probability $\frac{m_{S_1}}{q(i)}$, and the packet is successfully received at the destination with probability $\mu_{id}$.  Thus, packets from $S_1$ will leave the system with rate
\begin{equation}
\sum_{i\in S_1} \frac {m_{S_1}}{q(i)} \mu_{id}
\end{equation}

The final possible transition type occurs when a packet located at the nodes in subset $S_1$ is successfully received at some set of nodes which did not previously have that packet, but not the destination $d$, resulting in that packet being finally in the subset $S_2\supset S_1$.  Thus $m_{S_1}$ will decrease by $1$, and $m_{S_2}$ will increase by $1$.  The only terms $V_S(\underline{m})$ that will change are those containing $m_{S_1}$ but not $m_{S_2}$.  Thus, for $S$ such that $S\supseteq S_1$ and $S\nsupseteq S_2$,
\begin{align} \notag  
V_S&(m_0,m_1,...,m_{S_1}-1,...,m_{S_2}+1,...)-V_S(m_0,m_1,...)
\\ \notag
&=N_{|S|}\left(1+\delta\right)^{\sum_{S'\subseteq S} m_{S'}-1}-
N_{|S|}\left(1+\delta\right)^{\sum_{S'\subseteq S} m_{S'}}\\
&=-\delta N_{|S|}\left(1+\delta\right)^{\sum_{S'\subseteq S} m_{S'}-1}.
\end{align}
These events occur when any node $i\in S_1$ transmits a packet in $S_1$, and that packet is successfully received by all the nodes $j\in S_2/S_1$, and not successful in reaching nodes $\{j|j\notin S_2\}$, including the destination node $d$.  The total rate of such events is 
\begin{equation}
\sum_{i\in S_1} \frac {m_{S_1}}{q(i)} \prod_{j\in S_2/S_1} \mu_{ij} \prod_{j\notin S_2} \epsilon_{ij}.
\end{equation}

The expected increase in the total Lyapunov function due to arrivals should be less than the expected decrease due to departures and transitions on all but a finite number of state $\underline{m}$.  The sum of changes over all of the terms must therefore satisfy
\begin{align} \notag
\lambda &\sum_{S\in\mathcal{S}}{N_{|S|}\left(1+\delta\right)^{\sum_{S'\subseteq S} m_{S'}}} \\
&<\sum_{S\in\mathcal{S}} \sum_{\{\left(S_1,S_2\right)|S_1 \subset S_2,S_1 \subseteq S,S_2 \nsubseteq S\}}
\left(\sum_{i\in S_1} \frac {m_{S_1}}{q(i)} \prod_{j\in S_2/S_1} \mu_{ij} \prod_{j\notin S_2} \epsilon_{ij}\right)
N_{|S|}\left(1+\delta\right)^{\sum_{S'\subseteq S} m_{S'}-1} \notag \\
&+ \sum _{S\in\mathcal{S}} \sum_{S_1\subseteq S} \left( \sum_{i\in S_1} \frac {m_{S_1}}{q(i)} \mu_{id}\right)
N_{|S|}\left(1+\delta\right)^{\sum_{S'\subseteq S} m_{S'}-1} \label{eq:big}
\end{align}

In the second line of Equation (\ref{eq:big}), the first summation is over terms in the Lyapunov function.  The second summation is over transitions of the possible pairs of $S_1$ and $S_2$ which will effect that particular term, and the third summation is over nodes which could possibly transmit and create that transition.  The final terms of the second line represent the value of the change in that term $V_S(\underline{m})$.

Similarly, in the third line of Equation (\ref{eq:big}), the first summation is over the terms of the Lyapunov function, and the second is over the possible departures from the system which can effect the value of each term.  Within the parentheses is the rate of those departures, and the final terms again represent the value of the change in the term $V_S(\underline{m})$.

Note that if $q(i)=0$, that is, no packets are currently in the queue at node $i$, then for any $S$ such that $i\in S$, $m_S=0$.   In this case, take
\begin{equation*}
\frac{m_S}{q(i)}=\frac{0}{0}=0
\end{equation*} 
since this node cannot transmit any packets.

Combining the two terms on the righthand side of Equation (\ref{eq:big}) yields
\begin{align} \label{eq:s1}
\sum_{S\in\mathcal{S}} \sum_{S_1\subseteq S} \sum_{i\in S_1} \frac {m_{S_1}}{q(i)} \left( \mu_{id}+\epsilon_{id} \sum_{\{S_2|S_1\subset S_2 ,S_2 \nsubseteq S\}}      \prod_{j\in S_2/S_1} \mu_{ij} \prod_{j\notin S_2,j\ne d} \epsilon_{ij}    \right)
N_{|S|}\left(1+\delta\right)^{\sum_{S'\subseteq S} m_{S'}-1}.
\end{align}

We can simplify Equation (\ref{eq:s1}) by observing the fact that if $A=\{1,2,...,n\}$ and $0\le p_j \le 1$ then
\begin{align} \label{eq:1sum}
1=\sum_{A_1 \subseteq A} \left( \prod_{j\in A_1} p_j \prod_{j\notin A_1} \left(1-p_j \right) \right).
\end{align}
Equation (\ref{eq:1sum}) can be proven by induction on the size of $A$, or by interpreting the $p_j$ as the probabilities that each of $n$ independent events occur.  Each product term is the probability that exactly the subset $A_1$ of the events occurs, and so the sum over all $A_1$'s is $1$.

Fix some subset $S$, another subset $S_1 \subseteq S$, and any $i\in S$. Then, we have 
\begin{align} \label{eq:sum_sequence1}
1&=\sum_{\{S_2|S_2\supseteq S_1\}} \left (\prod_{j\in S_2/S_1} \mu_{ij} \prod_{j\notin S_2,j\ne d} \epsilon_{ij}\right)\\ \label{eq:sum_sequence2}
1&=\sum_{\{S_2|S_2\supseteq S_1,S_2\subseteq S\}} \left(\prod_{j\in S_2/S_1} \mu_{ij} \prod_{j\notin S_2,j\ne d} \epsilon_{ij}\right)
+  \sum_{\{S_2|S_2\supset S_1,S_2 \nsubseteq S\}} \left(\prod_{j\in S_2/S_1} \mu_{ij} \prod_{j\notin S_2,j\ne d} \epsilon_{ij} \right) \\ \label{eq:sum_sequence3}
1&=\prod_{j\notin S,j\ne d}\epsilon_{ij} \left[ \sum_{\{S_2|S_2\supseteq S_1,S_2\subseteq S\}} \left(\prod_{j\in S_2/S_1} \mu_{ij} \prod_{j\in S/S_2} \epsilon_{ij}\right) \right]
+  \sum_{\{S_2|S_2\supset S_1,S_2 \nsubseteq S\}} \left(\prod_{j\in S_2/S_1} \mu_{ij} \prod_{j\notin S_2,j\ne d} \epsilon_{ij} \right)\\ \label{eq:sum_sequence4}
&\sum_{\{S_2|S_2\supset S_1,S_2 \nsubseteq S\}} \left(\prod_{j\in S_2/S_1} \mu_{ij} \prod_{j\notin S_2,j\ne d} \epsilon_{ij} \right)= 1-\prod_{j\notin S,j\ne d} \epsilon_{ij}
\end{align}
where Equation (\ref{eq:sum_sequence1}) is obtained as follows:  for each $S_2\supseteq S_1$, treat the $S_2/S_1$ as the $A_1$ in (\ref{eq:1sum}).  Equation  (\ref{eq:sum_sequence2}) splits the summation, (\ref{eq:sum_sequence3}) factors out some $\epsilon_{ij}$, and (\ref{eq:sum_sequence4}) recognizes that the term in brackets in (\ref{eq:sum_sequence3}) is again equal to unity by (\ref{eq:1sum}).

Combining Equations (\ref{eq:big}), (\ref{eq:s1}), (\ref{eq:sum_sequence4}), and recalling the definition of $C_i(S)$ from Equation (\ref{eq:ci}) yields the requirement
\begin{align} \notag
\lambda &\sum_{S\in\mathcal{S}}{N_{|S|}\left(1+\delta\right)^{\sum_{S'\subseteq S} m_{S'}}} \\ \label{eq:simplebig}
&<\sum_{S\in\mathcal{S}} \sum_{S_1\subseteq S} \sum_{i\in S_1} \left(\frac {m_{S_1}}{q(i)} C_i(S) \right)
N_{|S|}\left(1+\delta\right)^{\sum_{S'\subseteq S} m_{S'}-1}.
\end{align}

We must show that Equation (\ref{eq:simplebig}) holds for all but a finite number of states $\underline{m}$.  To begin, consider the states of the form $\underline{m}=(0,0,...,0,m_{S''},0,...,0)$, where all but a single one of the $2^n$ variables $m_{S}=0$.  As in Section \ref{sec:n1}, each of these states will provide the individual cut-set bounds on $\lambda$ required for stability by the theorem.  For $\underline{m}$ of this form, Equation (\ref{eq:simplebig}) reduces to
\begin{align} \label{eq:1nz}
\lambda &\sum_{S\supseteq S''} N_{|S|} \left(1+\delta\right)^{m_{S''}}
+\lambda \sum_{S\nsupseteq S''} N_{|S|} \\ \notag
&< \sum_{S\supseteq S''} \left(\sum_{i\in S''} C_i(S)\right) N_{|S|} \left(1+\delta\right)^{m_{S''}-1}
\end{align}
since $m_{S_1}=0$ for all $S_1\ne S''$ and $m_{S''}=q(i)$ when $i\in S''$.
As long as  $N_{|S''|}$ is chosen such that
\begin{equation}\notag
N_{|S''|} + \sum_{S\supset S''} N_{|S|} < N_{|S''|}\frac{N+1}{N}
\end{equation}
which is equivalent to the requirement of Equation (\ref{eq:define_n}), we can replace Equation (\ref{eq:1nz}) with
\begin{align} 
\lambda N_{|S''|}&\frac{N+1}{N}(1+\delta)
 +\lambda \sum_{S\nsupseteq S''} N_{|S|} \left(1+\delta\right)^{-m_{S''}+1}  
<C(S'') N_{|S''|}.\label{eq:1nz2}
\end{align}
The left-hand side has been increased with the substitution.  The right-hand side has been decreased, because all of the terms are positive, and instead of summing over all $S\in S''$ in the outer summation, we include only the term where $S=S''$.  Therefore, satisfying Equation (\ref{eq:1nz2}) assures that Equation (\ref{eq:1nz}) holds.

Choose $m_{S''}$ such that
\begin{equation} \notag
\frac{1}{N_{|S''|}} \sum_{S\nsupseteq S''} N_{|S|} \left(1+\delta\right)^{-m_{S''}+1} < \frac{N+1}{N}(1+\delta)^2 - \frac{N+1}{N}(1+\delta)
\end{equation}
and  Equation (\ref{eq:1nz2}) reduces to $\lambda < \frac{N}{N+1}\frac{1}{(1+\delta)^2} C(S'')$.

Thus, it has been shown that for states of the form $\underline{m}=(0,0,...,0,m_{S''},0,...,0)$ whenever $\lambda < \frac{N}{N+1}\frac{1}{(1+\delta)^2} C(S'')$, there exists a $\tilde{m}_{S''}$ sufficiently large such that the expected value of the Lyapunov function is decreasing for $m_{S''}>\tilde{m}_{S''}$.  Each of the required cut-set bounds on $\lambda$ for all of the different cuts $S\in\mathcal{S}$ are obtained in this manner.

It remains to show that there are only a finite number of general $\underline{m}=(m_0,m_1,...,m_{S},...,m_{2^n-1})$ where the expected value of the Lyapunov function is increasing.  That is, we will demonstrate that the cutset bounds on $\lambda$ obtained above are sufficient to guarantee that Equation (\ref{eq:L}) is indeed a Lyapunov function for the queuing system.

To do so, first examine the state variable $m_{2^n-1}$; that is,  the variable counting the number of packets which appear at every node in the system other than the destination.  We will show that there exists a finite $m_{2^n-1}^*$ for which, as long as $m_{2^n-1}> m_{2^n-1}^*$, regardless of the value of $m_0,m_1$, and every other state variable up to $m_{2^n-2}$, the expected value of the Lyapunov function Equation (\ref{eq:simplebig}) will be decreasing.

Let $\hat{S}$ be the subset $\hat{S}\in\mathcal{S}$ which contains all $n$ relay nodes and the source node $s$, i.e. the largest subset of the nodes.  Also, let $N_{|\hat{S}|}=1$.  Equation (\ref{eq:simplebig}) can be rewritten as
\begin{align} \notag
\lambda &\left(1+\delta\right)^{m_{\hat{S}}+\sum_{S'\subset\hat{S}} m_{S'}} 
+ \lambda \sum_{S\subset \hat{S}} N_{|S|} \left(1+\delta\right)^{\sum_{S'\subseteq S} m_{S'}} \\
\notag
&< \sum_{S_1\subseteq \hat{S}} \left (\sum_{i\in S_1} \frac{m_{S_1}}{q(i)} C_i(\hat{S})\right)
\left(1+\delta\right)^{m_{\hat{S}}+\sum_{S'\subset\hat{S}} m_{S'}-1} \\ \label{eq:lastterm}
& +\sum_{S\subset \hat{S}}  
\sum_{S_1\subseteq S} \left( \sum_{i\in S_1} \frac {m_{S_1}}{q(i)} C_i(\hat{S})\right)
N_{|S|}\left(1+\delta\right)^{\sum_{S'\subseteq S} m_{S'}-1}
\end{align}
By dividing Equation (\ref{eq:lastterm}) through by 
\begin{equation*}
\left(1+\delta\right)^{m_{\hat{S}}+\sum_{S'\subset\hat{S}} m_{S'}-1},
\end{equation*}
increasing the second term on the left-hand side, and decreasing the right-hand side by dropping the final term, the constraint
\begin{align} \label{eq:s_hat}
\lambda &(1+\delta) +\lambda   (1+\delta)^{1-m_{\hat{S}}} \sum_{S\subset \hat{S}} N_{|S|}  \\ \notag
& < \sum_{S_1\subseteq \hat{S}} \sum_{i\in S_1} \frac{m_{S_1}}{q(i)} C_i(\hat{S}) \\
& = \sum_{i\in \hat{S}} \left(\sum_{\{S_1|i\in S_1\}} \frac{m_{S_1}}{q(i)}\right) C_i(\hat{S})
=\sum_{i\in\hat{S}} C_i(\hat{S}) = C(\hat{S})
\end{align}
is obtained.

Therefore there exists a $m_{\hat{S}}^*$ for which, for all states $\underline{m}$ with $m_{\hat{S}}>m_{\hat{S}}^*$, regardless of the values of the other state variables, the expected value of the Lyapunov function will be decreasing when $\lambda < C(\hat{S})\frac{1}{(1+\delta)^2}$.

Next, consider any set $\hat{\hat S}$ which contains all but one of the $n$ relay nodes.  
Rearrange the summations in Equation (\ref{eq:simplebig}) to obtain the sufficient condition
\begin{align} \notag
\lambda &\sum_{S\in\mathcal{S}}{N_{|S|}\left(1+\delta\right)^{\sum_{S'\subseteq S} m_{S'}}} \\
&<\sum_{S\in\mathcal{S}} \left( \sum_{i\in S} \frac{\sum_{\{S_1|S_1\subseteq S,i\in S_1\}} m_{S_1}}{q(i)}   C_i(S) \right)
N_{|S|}\left(1+\delta\right)^{\sum_{S'\subseteq S} m_{S'}-1}
\label{eq:simplebig2}
\end{align}

Assume that $m_{\hat S} < m_{\hat{S}}^*$.  If  $m_{\hat{\hat S}}> m_{\hat{S}}^* \frac{2^n}{\delta}$, it will be shown that Equation (\ref{eq:simplebig2}) holds for any $\lambda < \frac{N}{N+1}\frac{1}{\left(1+\delta\right)^2} \min_{S\in\mathcal{S}}C(S)$.

There are two cases to consider:
\begin{itemize}
\item None of the $m_S$ other than $m_{\hat{\hat S}}$ are greater than $m_{\hat{\hat S}} \frac{\delta}{2^n}$.
\item At least one of the $m_S$ is greater than $m_{\hat{\hat S}}\frac{\delta}{2^n} $.
\end{itemize}

In the first case, divide all the sets $S$ into two classes: $\mathcal{S}_1 = \{S|\hat{\hat S}\subseteq S\}$ and $\mathcal{S}_1^C$. For $S\in\mathcal{S}_1$,
\begin{align*}
\frac{\sum_{\{S_1|S_1\subseteq S,i\in S_1\}} m_{S_1}}{q(i)} > \frac{m_{S''}}{m_{S''}+\delta m_{S''}} = \frac{1}{1+\delta},
\end{align*}
so term by term of the outer summation of Equation (\ref{eq:simplebig2}), when $\lambda < \frac{N}{N+1}\frac{1}{\left(1+\delta\right)^2} C(S)$, the terms in the class $\mathcal{S}_1$ are satisfied. 

Now consider any of the terms in the class $S\in\mathcal{S}_1^C$.  Divide both sides of Equation (\ref{eq:simplebig2}) by 
\begin{align*}
m_{sum}=\left(1+\delta\right)^{\sum_{S'\subseteq \hat{\hat S}} m_{S'}-1}
\end{align*}
and note that $m_{sum}$ is greater than the sum in the exponent of the $(1+\delta)$ for any the terms $S\in \mathcal{S}_1^C$ by more than $m_{\hat S}^*$.  The contribution to the left hand side from these terms becomes arbitrarily small, just as in Equation (\ref{eq:s_hat}).  Equation (\ref{eq:simplebig2}) is thus satisfied when none of the $m_S$ other than $m_{\hat{\hat S}}$ are greater than $m_{\hat{\hat S}} \frac{1}{2^n}$.

In the case where at least one of the other $m_S$ is greater than $m_{\hat{\hat S}}\frac{\delta}{2^n}$, define
\begin{align*}
S_U = \bigcup_{\{S|m_S>m_{\hat{\hat S}}\frac{\delta}{2^n}\}} S
\end{align*}
as the union of all these sets $S$.  The same analysis holds from the above case:  For each term $S\in \mathcal S$ of Equation (\ref{eq:simplebig2}), either $S \supseteq S_U$ and the right hand side is greater than a $\frac{1}{1+\delta}$ fraction of $C(S)$, or the exponent of that term is more than $m_{\hat{S}}^*$ less than the exponent of the $S_U$ term, and is thus inconsequential.

Define $m_S^*$ so that whenever $|S_1| = |S_2|-1$, $m_{S_1}^*>m_{S_2}^*\frac{2^n}{\delta}$.  The same arguments already made are used inductively to show that as long as $m_S>m_S^*$ for at least one $S$, then Equation (\ref{eq:simplebig2}) is satisfied.
\end{proof}  

\section{Simulation Results}
The proposed transmission algorithm was simulated for a proof-of-concept.  This section contains a description of the simulation methods and a presentation of the results.

The simulation was preformed in a slotted-time model:  At each time-step, a new packet arrives at the source node with a probability $\lambda$.  The simulation then loops over each node in the network, choosing a packet randomly from that node's queue, determining at which receivers the packet is successfully received, and adjusting the remaining queue states accordingly.  

\begin{figure}
\centering
\includegraphics[scale=0.6]{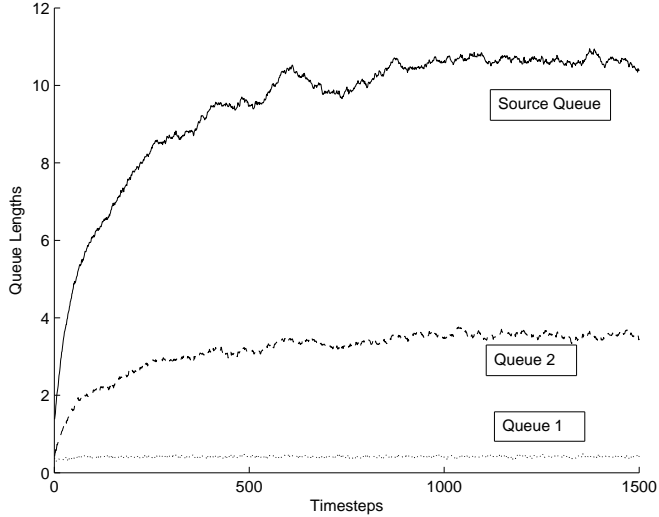}
\caption{Queue Lengths Averaged over 500 Trials} \label{fig:simresults}
\end{figure}

The simulation was run on the network of Figure \ref{fig:ex_2node}, with the erasure probabilities $\epsilon_{s1}=.6$, $\epsilon_{s2}=.5$, $\epsilon_{2d}=.9$, and $\epsilon_{1d}=.1$. The minimum cut of this network is the subset $S=\{s,2\}$, where the min-cut capacity is $.5$ packets/timeslot.  The source node received new packets at an arrival rate of $\lambda = .45$.  Figure \ref{fig:simresults} plots the queue lengths for a simulation of 1500 timesteps, averaged over 500 trial runs of the simulation.  

The source node and relay node $2$, which are to the left of the minimum cut, have relatively longer average lengths than relay node $1$, which is on the right side of the cutset.  Additional simulations were performed for all three of the possible minimum-cut configurations of this network (by adjusting the erasure probabilities), and for various other simple network configurations.

\section {Model Discussion and Extensions}
The wireless erasure network with feedback model we study is designed to take into account the most salient characteristics of a wireless network.  Namely, we model the wireless broadcast constraint on our directed graph, and consider the random erasures to be a model of an on/off type of fading.  This model acts as something between a strictly physical layer model and a higher-level abstraction such as the network layer.  For example, we assume that communication is packet-based and that the number of bits per packet, the error-protection level within a packet, and the modulation scheme are fixed.  For analytical reasons, we assume that the feedback is instantaneous and perfect, that erasure events are all independent across both time and space, and that packets are received without interference.  To justify the lack of interference we note, as in \cite{DGH:06}, that we assume some sort of interference-avoidance is already built into the network.  Additionally, we conjecture that in a slotted-time model, correlation of error events at each receiver can be dealt with similarly as it is in \cite{DGH:06}.  Feedback delay and spatial correlation of erasure events are further discussed in this section.

\subsection{Non-Instantaneous and Imperfect Feedback}
Consider the case where, instead of immediately being available at all nodes, the feedback suffers from a delay which is exponentially distributed with mean $D$.  The model of this paper can easily accommodate this extension:  after the actual destination node of the network, insert at least $D$ loss-free ($\epsilon_{ij}=0$) links in series.  Create a new, virtual destination node at the end of this sequence of links.  Adding in the new links does not decrease the cut-set bound, so the new network remains stable under the given transmission strategy.  The new network acts precisely as the old network,  with the inclusion of a delayed feedback that is at least as bad as exponentially distributed with mean $D$, performs.

If the feedback is to contain errors, that it, if some of the relay nodes mistakenly do not receive feedback, this can be taken care of by allowing some link level feedback:  The network can include a mechanism by which, if a node receives a packet which that node has already determined should be flushed, it can send a single link-level feedback to the offending relay.  Again, the process reduces simply to a network with feedback delay.  

\subsection {Spatial Correlation of Dropped Packets}
It is also possible to consider a model the dropping of packets transmitted from each transmitter $i\in V$ are correlated events.  That is, if the node $i$ transmits a packet, the probability that exactly the set $W\subseteq V$ successfully receives that packet can be considered to be $p(i,W)$.  In the independent model used in the majority of this paper,
\begin{align*}
p(i,W) = \prod_{j\in W} \mu_{ij} \prod_{j\in W^C, j\ne i} \epsilon_{ij},
\end{align*}
and 
\begin{align*}
\sum_{W \in \mathcal{S}} p(i,W) = 1
\end{align*}   
just as observed in Equation (\ref{eq:1sum}).

The cut-set bound for the $S-S^C$ cut can still be interpreted as the sum of rates for which nodes in $S$ can transmit and at least one node in $S^C$ will successfully receive the packet:
\begin{align}
C(S) = \sum_{i\in S} \sum_{\left\{W|W\cap S^C \ne \emptyset\right\}} p(i,W).
\end{align}
Replacing the transition probabilities in the Markov chain model with these $p(i,W)$ requires no substantive changes in the proof technique - the values of the cutset bounds $C(S)$ and the probabilities $C_i(S)$ change accordingly, and the proof of queue stability follows.

Allowing correlations across time for a single or multiple edges is a much different problem.  An entire new set layers of the Markov chain would be required, and the whether the cut-set bound is achievable is still unknown in even the feedback-free model of \cite{DGH:06}.  Because of the asynchronous nature of our model, this same difficulty is encountered if it is desired to correlate erasures of packets from different transmitters.  Such events would be simultaneous in the slotted-time model, and therefore are dealt with in \cite{DGH:06}, but would induce correlations over time in the model of this paper.

\section{Conclusion}
In this paper, we have demonstrated a parallel between the erasure channel and a network of such channels:  When acknowledgment feedback is available, there exists a simple transmission strategy by which the information-theoretic capacity (calculated by the cut-set bound) can be achieved for a unicast network without any need for a coding scheme, and without any knowledge of the erasure probabilities.  We have described a novel randomized and decentralized strategy which requires only a surprisingly small amount of information about the network (specifically, no knowledge whatsoever about the network topology) to succeed in stabilizing the queues and achieving throughput optimality.  The main results shows a tradeoff between network coding and feedback - given one, the the other is not required to design a throughput optimal algorithm.

While our randomized algorithm is throughput optimal, it will suffer in the metric of average packet delay.  If we desire to transmit $N$ packets through the network, for example, the total time required will be $N/C + o(N)$, which is optimal, but the average delay for a given packet may be $\Theta(N)$ because of the randomization.  Improving the delay performance will undoubtedly require more coordination and feedback among the nodes in the network and is worthy of further scrutiny, but beyond the scope of the current paper.


\end{document}

%% file: 2node.pstex_t
\begin{picture}(0,0)%
\includegraphics{2node.pstex}%
\end{picture}%
\setlength{\unitlength}{3947sp}%
\begingroup\makeatletter\ifx\SetFigFont\undefined%
\gdef\SetFigFont#1#2#3#4#5{%
  \reset@font\fontsize{#1}{#2pt}%
  \fontfamily{#3}\fontseries{#4}\fontshape{#5}%
  \selectfont}%
\fi\endgroup%
\begin{picture}(5463,1632)(-449,-3485)
\put(3076,-2116){\makebox(0,0)[lb]{\smash{{\SetFigFont{12}{14.4}{\rmdefault}{\mddefault}{\updefault}{\color[rgb]{0,0,0}$q(2)=m_{10}+m_{11}$}%
}}}}
\put(3076,-3316){\makebox(0,0)[lb]{\smash{{\SetFigFont{12}{14.4}{\rmdefault}{\mddefault}{\updefault}{\color[rgb]{0,0,0}$q(1)=m_{01}+m_{11}$}%
}}}}
\put(3601,-2686){\makebox(0,0)[lb]{\smash{{\SetFigFont{12}{14.4}{\rmdefault}{\mddefault}{\updefault}{\color[rgb]{0,0,0}$d$}%
}}}}
\put(2626,-2236){\makebox(0,0)[lb]{\smash{{\SetFigFont{12}{14.4}{\rmdefault}{\mddefault}{\updefault}{\color[rgb]{0,0,0}$2$}%
}}}}
\put(2626,-3436){\makebox(0,0)[lb]{\smash{{\SetFigFont{12}{14.4}{\rmdefault}{\mddefault}{\updefault}{\color[rgb]{0,0,0}$1$}%
}}}}
\put(-449,-3061){\makebox(0,0)[lb]{\smash{{\SetFigFont{12}{14.4}{\rmdefault}{\mddefault}{\updefault}{\color[rgb]{0,0,0}$q(s)=m_{00}+m_{01}+m_{10}+m_{11}$}%
}}}}
\put(1576,-2611){\makebox(0,0)[lb]{\smash{{\SetFigFont{12}{14.4}{\rmdefault}{\mddefault}{\updefault}{\color[rgb]{0,0,0}$s$}%
}}}}
\end{picture}%

%% file: 1node.pstex_t
\begin{picture}(0,0)%
\includegraphics{1node.pstex}%
\end{picture}%
\setlength{\unitlength}{3947sp}%
\begingroup\makeatletter\ifx\SetFigFont\undefined%
\gdef\SetFigFont#1#2#3#4#5{%
  \reset@font\fontsize{#1}{#2pt}%
  \fontfamily{#3}\fontseries{#4}\fontshape{#5}%
  \selectfont}%
\fi\endgroup%
\begin{picture}(3053,1420)(976,-3074)
\put(3601,-2686){\makebox(0,0)[lb]{\smash{{\SetFigFont{12}{14.4}{\rmdefault}{\mddefault}{\updefault}{\color[rgb]{0,0,0}$d$}%
}}}}
\put(1951,-2236){\makebox(0,0)[lb]{\smash{{\SetFigFont{12}{14.4}{\rmdefault}{\mddefault}{\updefault}{\color[rgb]{0,0,0}$\epsilon_{s1}$}%
}}}}
\put(3151,-2161){\makebox(0,0)[lb]{\smash{{\SetFigFont{12}{14.4}{\rmdefault}{\mddefault}{\updefault}{\color[rgb]{0,0,0}$\epsilon_{1d}$}%
}}}}
\put(2551,-2836){\makebox(0,0)[lb]{\smash{{\SetFigFont{12}{14.4}{\rmdefault}{\mddefault}{\updefault}{\color[rgb]{0,0,0}$\epsilon_{sd}$}%
}}}}
\put(976,-3016){\makebox(0,0)[lb]{\smash{{\SetFigFont{12}{14.4}{\rmdefault}{\mddefault}{\updefault}{\color[rgb]{0,0,0}$q(s)=m_0+m_1$}%
}}}}
\put(3076,-1816){\makebox(0,0)[lb]{\smash{{\SetFigFont{12}{14.4}{\rmdefault}{\mddefault}{\updefault}{\color[rgb]{0,0,0}$q(1)=m_1$}%
}}}}
\put(3601,-2686){\makebox(0,0)[lb]{\smash{{\SetFigFont{12}{14.4}{\rmdefault}{\mddefault}{\updefault}{\color[rgb]{0,0,0}$d$}%
}}}}
\put(3601,-2686){\makebox(0,0)[lb]{\smash{{\SetFigFont{12}{14.4}{\rmdefault}{\mddefault}{\updefault}{\color[rgb]{0,0,0}$d$}%
}}}}
\put(3601,-2686){\makebox(0,0)[lb]{\smash{{\SetFigFont{12}{14.4}{\rmdefault}{\mddefault}{\updefault}{\color[rgb]{0,0,0}$d$}%
}}}}
\put(1576,-2686){\makebox(0,0)[lb]{\smash{{\SetFigFont{12}{14.4}{\rmdefault}{\mddefault}{\updefault}{\color[rgb]{0,0,0}$s$}%
}}}}
\put(2626,-1786){\makebox(0,0)[lb]{\smash{{\SetFigFont{12}{14.4}{\rmdefault}{\mddefault}{\updefault}{\color[rgb]{0,0,0}$1$}%
}}}}
\end{picture}%

%% file: 1node_chain.pstex_t
\begin{picture}(0,0)%
\includegraphics{1node_chain.pstex}%
\end{picture}%
\setlength{\unitlength}{3947sp}%
\begingroup\makeatletter\ifx\SetFigFont\undefined%
\gdef\SetFigFont#1#2#3#4#5{%
  \reset@font\fontsize{#1}{#2pt}%
  \fontfamily{#3}\fontseries{#4}\fontshape{#5}%
  \selectfont}%
\fi\endgroup%
\begin{picture}(4958,2473)(297,-3960)
\put(2217,-3661){\makebox(0,0)[lb]{\smash{{\SetFigFont{12}{14.4}{\rmdefault}{\mddefault}{\updefault}{\color[rgb]{0,0,0}$(m_0,m_1-1)$}%
}}}}
\put(3976,-2686){\makebox(0,0)[lb]{\smash{{\SetFigFont{12}{14.4}{\rmdefault}{\mddefault}{\updefault}{\color[rgb]{0,0,0}$(m_0+1,m_1)$}%
}}}}
\put(301,-1891){\makebox(0,0)[lb]{\smash{{\SetFigFont{12}{14.4}{\rmdefault}{\mddefault}{\updefault}{\color[rgb]{0,0,0}$(m_0-1,m_1+1)$}%
}}}}
\put(376,-2716){\makebox(0,0)[lb]{\smash{{\SetFigFont{12}{14.4}{\rmdefault}{\mddefault}{\updefault}{\color[rgb]{0,0,0}$(m_0-1,m_1)$}%
}}}}
\put(3376,-2461){\makebox(0,0)[lb]{\smash{{\SetFigFont{12}{14.4}{\rmdefault}{\mddefault}{\updefault}{\color[rgb]{0,0,0}$\lambda$}%
}}}}
\put(2326,-2686){\makebox(0,0)[lb]{\smash{{\SetFigFont{12}{14.4}{\rmdefault}{\mddefault}{\updefault}{\color[rgb]{0,0,0}$(m_0,m_1)$}%
}}}}
\put(1876,-2086){\makebox(0,0)[lb]{\smash{{\SetFigFont{12}{14.4}{\rmdefault}{\mddefault}{\updefault}{\color[rgb]{0,0,0}$\mu_{s1}\epsilon_{sd}\frac{m_0}{m_0+m_1}$}%
}}}}
\put(2851,-3136){\makebox(0,0)[lb]{\smash{{\SetFigFont{12}{14.4}{\rmdefault}{\mddefault}{\updefault}{\color[rgb]{0,0,0}$\mu_{1d}+\mu_{sd}\frac{m_1}{m_0+m_1}$}%
}}}}
\put(1201,-3136){\makebox(0,0)[lb]{\smash{{\SetFigFont{12}{14.4}{\rmdefault}{\mddefault}{\updefault}{\color[rgb]{0,0,0}$\mu_{sd} \frac{m_0}{m_0+m_1}$}%
}}}}
\end{picture}%